\documentclass[aps,prl,a4paper,,showpacs,twocolumn,floatfix]{revtex4}
\usepackage[latin1]{inputenc}
\usepackage{amsfonts}
\usepackage{amssymb}
\usepackage{amsmath}
\usepackage{calc}
\usepackage{graphicx}
\usepackage{epsfig}
\usepackage{bm}
\usepackage{ulem}
\usepackage{setspace}
\usepackage{subfigure}

\begin{document}

\title{Purification of single-photon entanglement with linear optics}
\date{\today}
\author{Nicolas Sangouard,$^{1}$\footnotemark[1] \footnotetext{\footnotemark[1] These authors contributed equally
to this work.}  Christoph Simon,$^{2}$\footnotemark[1]
Thomas Coudreau,$^{1}$ and Nicolas Gisin$^{2}$}
\affiliation{%
$^{1}$Laboratoire MPQ, UMR CNRS 7162, Universit\'e Paris 7, France,\\
$^{2}$Group of Applied Physics, University of Geneva, Switzerland}

\begin{abstract}
We show that single-photon entangled states of the form
$|0\rangle|1\rangle+|1\rangle|0\rangle$ can be purified
with a simple linear-optics based protocol, which is
eminently feasible with current technology. Besides its
conceptual interest, this result is relevant for attractive
quantum repeater protocols.
\end{abstract}

\maketitle

Single-particle entanglement of the form
$|1\rangle_A|0\rangle_B+|0\rangle_A |1\rangle_B$ may be the
simplest form of entanglement \cite{VanEnk05}. It
corresponds to a single particle that is in a superposition
state of being in location (mode) A and of being in
location B. The particle can be a freely propagating photon
as in proposals for linear-optics quantum computing
\cite{KLM01}, but also e.g. a single excitation in an
atomic-ensemble based quantum memory, i.e. a stored photon
\cite{Chou05}. Single-photon entanglement has been
teleported experimentally in a purely photonic experiment
\cite{Lombardi02}. Single-excitation entanglement in atomic
ensembles has also been used to implement the basic segment
of a quantum repeater \cite{Chou07}. Note that in principle
single-photon states of the form
$|1\rangle_A|0\rangle_B+|0\rangle_A |1\rangle_B$ can
furthermore be converted into two-atom entangled states of
the form $|e\rangle_A|g\rangle_B+|g\rangle_A |e\rangle_B$,
cf. \cite{VanEnk05}, which are also often used in quantum
communication schemes \cite{eg}.

Entanglement purification is an important concept in
quantum information. It was introduced in
Ref.\cite{Bennett96}, which showed that it is possible to
convert two copies of a less entangled state into one copy
of a more entangled state using only local operations and
classical communication. The first entanglement
purification protocols \cite{Bennett96,Deutsch96} were
formulated in terms of qubits and quantum gates, in
particular they required CNOT gates. Both for practical
applications and from a conceptual point of view it is of
great interest to look for implementations that are as
simple as possible. For example, Ref. \cite{Pan01} proposed
a method for the purification of polarization-entangled
photon pairs that could be realized with linear optical
elements (without the need for CNOT gates). This procedure
was adapted to parametric down-conversion sources in Ref.
\cite{Simon02}, leading to an experimental realization
\cite{Pan03}.

An important domain of application for entanglement
purification is in the context of long-distance quantum
communication. The direct distribution of quantum states is
limited by the problem of photon loss in transmission. This
can be overcome using quantum repeater protocols
\cite{Briegel98} based on the creation and storage of
entanglement for moderate-distance elementary links,
followed by entanglement swapping. In practice these
operations cannot be performed with perfect fidelity,
limiting the number of links that can be used. This number
and thus the achievable distance can be greatly increased
if entanglement purification is used.

\begin{figure}
{\includegraphics[scale=0.32]{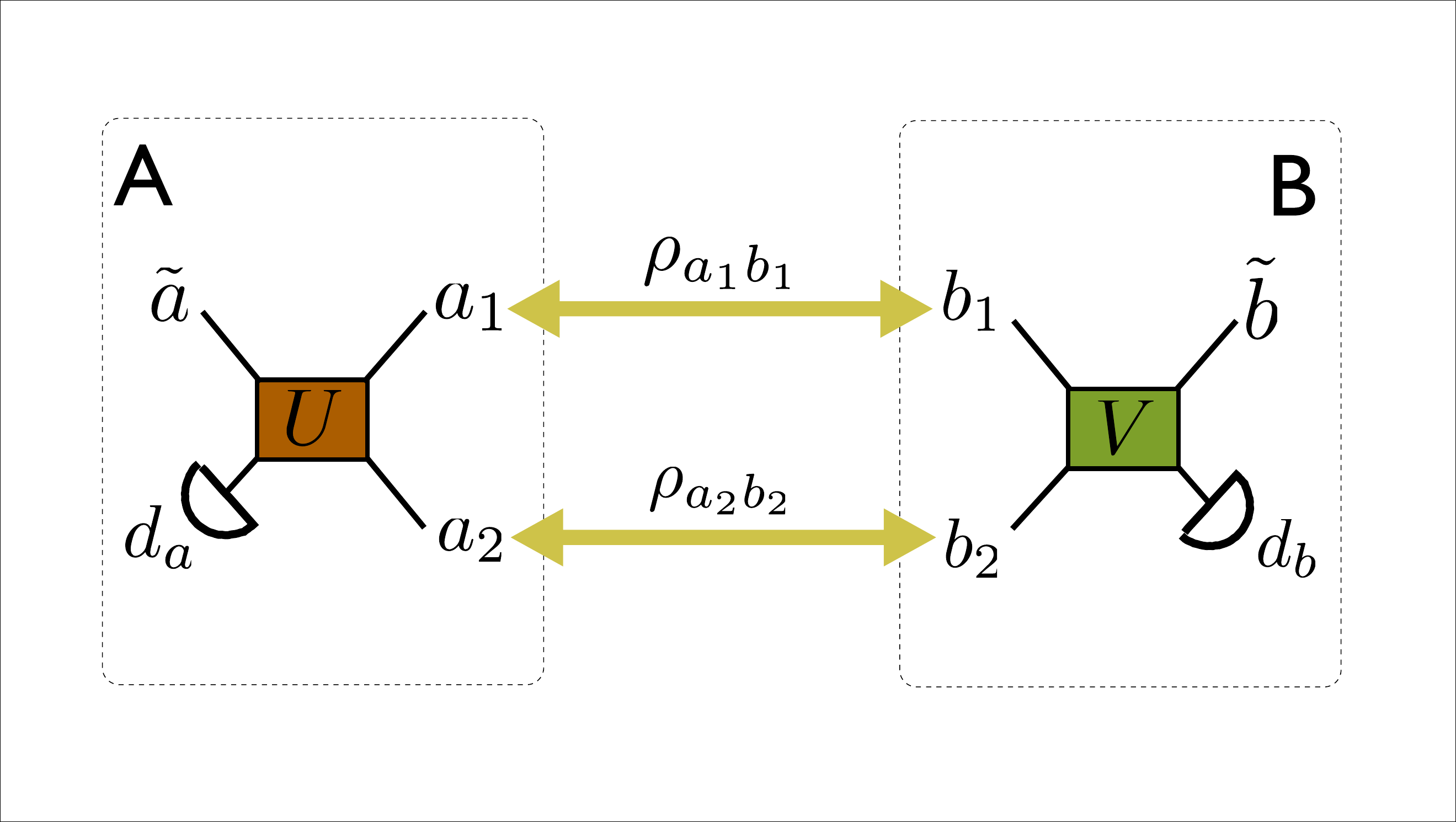} \caption{(Color
online) Scheme for entanglement purification of single-
photon entanglement. Alice and Bob share two entangled
single-photon states $\rho_{a_1b_1},$ $\rho_{a_2b_2}$ with
fidelity $F$. Both parties apply local unitary
linear-optical transformations $U$ and $V$ to their
respective modes. For appropriately chosen $U$ and $V$, the
detection of one photon in either $d_a$ or $d_b$ projects
modes $\tilde{a}$ and $\tilde{b}$ into a single-photon
entangled state with higher fidelity.}\label{fig1}}
\end{figure}

The linear-optics purification protocol of Ref.
\cite{Pan01} can be readily integrated into quantum
repeater protocols that are based on photon-pair
entanglement \cite{TwoPhotonRepeaters,SangouardIPP}.
However, there are other very attractive protocols based on
single-photon detections \cite{SinglePhotonRepeaters,
Sangouard07}, see e.g. Ref. \cite{Chou07} for a recent
related experiment. While being slower than the best known
protocol \cite{SangouardIPP}, they are rather simple and
require significantly fewer resources to outperform the
direct transmission of photons. They will thus be easier to
implement in the short term, and may well be the first
repeater protocols achieving a genuine advantage compared
to direct transmission. (More detailed resource counts are
given in Ref. \cite{SangouardIPP}.) As a consequence of
their relative simplicity, protocols based on single-photon
detections are also significantly less sensitive to
imperfections, such as non-unit memory read out or photon
detection efficiencies, as compared with protocols based on
two-photon detections. The main drawback of protocols based
on single-photon detections is that, unlike protocols based
on two-photon detections, they are interferometrically
sensitive to path length fluctuations  \cite{Minar07}.
Purification of single-photon entanglement is thus
particularly important in this context. However, to our
knowledge, no purification procedure for single-photon
entanglement has been proposed so far.

Here we show that single-photon entanglement purification
can be realized in a very simple way using only linear
optical elements and photon detectors. Suppose that Alice
and Bob ideally want to share a maximally entangled state
$\psi_+^{ab}=\frac{1}{\sqrt{2}}\left(|1\rangle_A
|0\rangle_B+|0\rangle_A |1\rangle_B
\right)=\frac{1}{\sqrt{2}}\left(a^{\dagger}+b^{\dagger}\right)|0\rangle$,
but that there are phase errors due e.g. to channel length
fluctuations, such that without entanglement purification
they can only distribute copies of the state
\begin{equation}
\label{initial_state}
\rho_{ab}=F|\psi_+^{ab}\rangle\langle\psi_+^{ab}|+(1-F)|\psi_-^{ab}\rangle\langle\psi_-^{ab}|,
\end{equation}
with $\frac{1}{2}<F<1$, where
$\psi_-^{ab}=\frac{1}{\sqrt{2}}(a^\dagger-b^\dagger)|0\rangle$
is the state after a phase error \cite{phase};
$F=\frac{1}{2}$ corresponds to the case where all phase
information has been lost and no entanglement is left. Note
that losses and phase errors are the most significant
practical limitations in the present context. In the
context of quantum repeaters, vacuum
($|0_A\rangle|0_B\rangle$) and multi-photon
($|1_A\rangle|1_B\rangle$) components are also relevant
errors. However, vacuum components do not decrease the
fidelity of the distributed state since the final
measurement in schemes based on single-photon detections
post-selects the cases where there was a photon in the
output. For multi-photon components, one can use a specific
architecture based on single-photon sources which does not
create multi-photon components in the elementary link,
making it very efficient \cite{Sangouard07}.

We now show how, starting from two copies of the state Eq.
(\ref{initial_state}), one can create one copy of a state
of the same form with fidelity $\tilde{F}>F$. We look for a
purification protocol that has the simple form shown in
Fig. 1. Alice and Bob both perform a linear unitary
transformation on their two modes ($a_1$ and $a_2$, and
$b_1$ and $b_2$ respectively), and then they each detect
one of the output modes ($d_a$ and $d_b$). The goal is to
have a higher-fidelity single-photon entangled state in
modes $\tilde{a}$ and $\tilde{b}$, cf. Fig. 1. The events
of interest are thus those where one photon is detected in
either $d_a$ or $d_b$, leaving the other photon in the
modes $\tilde{a},\tilde{b}$. The most general form for the
matrix describing Alice's transformation is
\begin{equation}
U(\theta,\phi,\xi)=\left[
\begin{matrix}
\cos\theta e^{i\xi} & -\sin\theta e^{-i\phi} \\
\sin\theta e^{i\phi} & \cos\theta e^{-i\xi},
\end{matrix}
\right]
\end{equation}
with $\{a_1^\dagger|0\rangle,
a_2^{\dagger}|0\rangle\}=U\{\tilde{a}^{\dagger}|0\rangle,d_a^{\dagger}|0\rangle\}.$
Bob can choose a different transformation, denoted
$V(\theta',\phi',\xi')$ which is obtained from $U$ by
replacing the arguments by $(\theta',\phi',\xi').$

We now determine the state of the output modes $\tilde{a}$
and $\tilde{b}$ conditional on the detection of a single
photon in mode $d_a$ and of zero photons in mode $d_b$. The
initial state $\rho_{a_1b_1}\otimes\rho_{a_2b_2}$ can be
seen as a probabilistic mixture of the four pure states
$\psi_{+}^{a_1b_1} \otimes \psi_{+}^{a_2b_2}$,
$\psi_{-}^{a_1b_1} \otimes \psi_{+}^{a_2b_2}$,
$\psi_{+}^{a_1b_1} \otimes \psi_{-}^{a_2b_2}$, and
$\psi_{-}^{a_1b_1} \otimes \psi_{-}^{a_2b_2}$ with
probabilities $F^2$, $F(1-F)$, $F(1-F)$, and $(1-F)^2$
respectively. These states give rise to conditional states
$\frac{1}{2}(A\tilde{a}^{\dagger} + B_-
\tilde{b}^{\dagger})|0\rangle$,
$\frac{1}{2}(A\tilde{a}^{\dagger} - B_+
\tilde{b}^{\dagger})|0\rangle$,
$\frac{1}{2}(A\tilde{a}^{\dagger} + B_+
\tilde{b}^{\dagger})|0\rangle$, and
$\frac{1}{2}(A\tilde{a}^{\dagger} - B_-
\tilde{b}^{\dagger})|0\rangle$ respectively, where
$A=\cos2\theta,$ and $B_\pm=\cos\theta e^{-i\xi}\cos\theta'
e^{i\xi'}\pm \sin\theta e^{-i\phi}\sin\theta' e^{i\phi'}$
are functions of the matrix elements of $U$ and $V$. Based
on these formulas one can calculate the trace of the
conditional state $\tilde{\rho}_{ab}$,
\begin{equation}
\label{trace_cond_state}
\hbox{tr}\tilde{\rho}_{ab}=\frac{A^2}{4}+\frac{F^2+(1-F)^2}{4}|B_-|^2+\frac{F(1-F)}{2}|B_+|^2,
\end{equation}
and its fidelity with respect to the desired state
$\psi_+^{ab}$,
\begin{equation}
\label{fidelity}
\tilde{F}=\frac{1}{2}+\frac{A(F^2-(1-F)^2)\mbox{Re}(B_-)}{A^2+(F^2+(1-F)^2)|B_-|^2+2F(1-F)|B_+|^2}.
\end{equation}
The case of one photon detected in $d_b$ and no photon in
$d_a$ leads to analogous expressions, it is sufficient to
exchange the roles of the matrix elements of $U$ and $V$
(i.e. primed and unprimed angles).

Our goal is to find the transformations $U$ and $V$ that
maximize the output fidelity Eq. (\ref{fidelity}). One can
see that the optimal fidelity is achieved for $B_+=0$ and
$\mbox{Im}(B_-)=0$. Both these conditions can be satisfied
choosing $\xi=\xi'=\phi=\phi'=0$ and
$\theta'=\theta-\frac{\pi}{2}$, giving $B_-=\sin 2\theta$.
Due to the above mentioned symmetry, this choice also
maximizes the output fidelity for a detection in $d_b$.
Note that the optimum choice of transformations $U$ and $V$
is asymmetric, i.e. $U \neq V$. The fidelity now depends on
the single parameter $\theta$,
\begin{equation}
\label{simplified_fidelity}
\tilde{F}=\frac{1}{2}+\frac{\left(F^2-\left(1-F\right)^2\right)\tan
2\theta}{1+\left(F^2+\left(1-F\right)^2\right)\tan^2
2\theta}.
\end{equation}

For situations where the fidelity of the initial states is
known, one can use Eq. (\ref{simplified_fidelity}) to
optimize the parameter $\theta$ as a function of $F$. One
finds
\begin{equation}
\label{optimal_angle} \tan
2\theta_{opt}=\frac{1}{\sqrt{F^2+(1-F)^2}},
\end{equation}
giving the optimized output fidelity
\begin{equation}
\label{fidelity_max}
\tilde{F}_{opt}=\frac{1}{2}+\frac{F^2-(1-F)^2}{2\sqrt{F^2+(1-F)^2}},
\end{equation}
cf. Fig. 2.

One can show that this is the optimal fidelity that can be
achieved even if one admits auxiliary input modes that are
initially empty. In this more general case the fidelity is
still given by Eq. (\ref{fidelity}), but $A$, $B_+$ and
$B_-$ now depend on the relevant coefficients of a larger
unitary matrix. However, optimizing Eq. (\ref{fidelity}) as
if $A$, $B_+$ and $B_-$ were independent variables, one
finds the same optimum expression Eq. (\ref{fidelity_max}).
This implies that Eq. (\ref{fidelity_max}) cannot be
improved for any number of auxiliary empty modes. The
situation may be different for auxiliary photons.

The success probability for the optimized protocol as
described above is $p_{opt}=2 \times \mbox{tr}
\tilde{\rho}_{ab}$ since both cases (detection in $d_a$, no
detection in $d_b$ and vice versa) contribute. One finds
\begin{equation}
p_{opt}=\frac{F^2+(1-F)^2}{1+F^2+(1-F)^2}.
\end{equation}
For situations where the input fidelity $F$ is not known a
priori, one has to choose a value of $\theta$ that is
independent of $F$. For example, if $F$ is unknown, but
expected to be close to 1, one could choose the value of
$\theta_{opt}$ for $F=1$, which gives
$\theta=\frac{\pi}{8}$. For this simplified protocol one
finds an output fidelity
\begin{equation}
\tilde{F}_{1}=\frac{F^2+F}{1+F^2+(1-F)^2}
\end{equation}
and a success probability
\begin{equation}
p_1=\frac{1+F^2+(1-F)^2}{4},
\end{equation}
which is close to 1/2 for $F$ close to 1. Fig. 2 shows the
output fidelities $\tilde{F}_{opt}$ and $\tilde{F}_1$ both for the optimal
protocol and for the simplified protocol. Both protocols
achieve substantial purification, and the simplified
protocol is almost as effective as the optimal one. It is
interesting to consider the regime of high fidelities. For
$F=1-\epsilon$ both $\tilde{F}_{opt}$ and $\tilde{F}_1$ are
equal to $1-\frac{\epsilon}{2}+O(\epsilon^2)$, i.e. the
purification protocol divides the error by a factor of 2.
Note that in quantum repeater protocols the error is
approximately doubled with every level of entanglement
swapping. This means that the present protocol has the
potential of significantly increasing the number of
possible levels, and thus the achievable total distance.

\begin{figure}
{\includegraphics[width=0.9 \columnwidth]{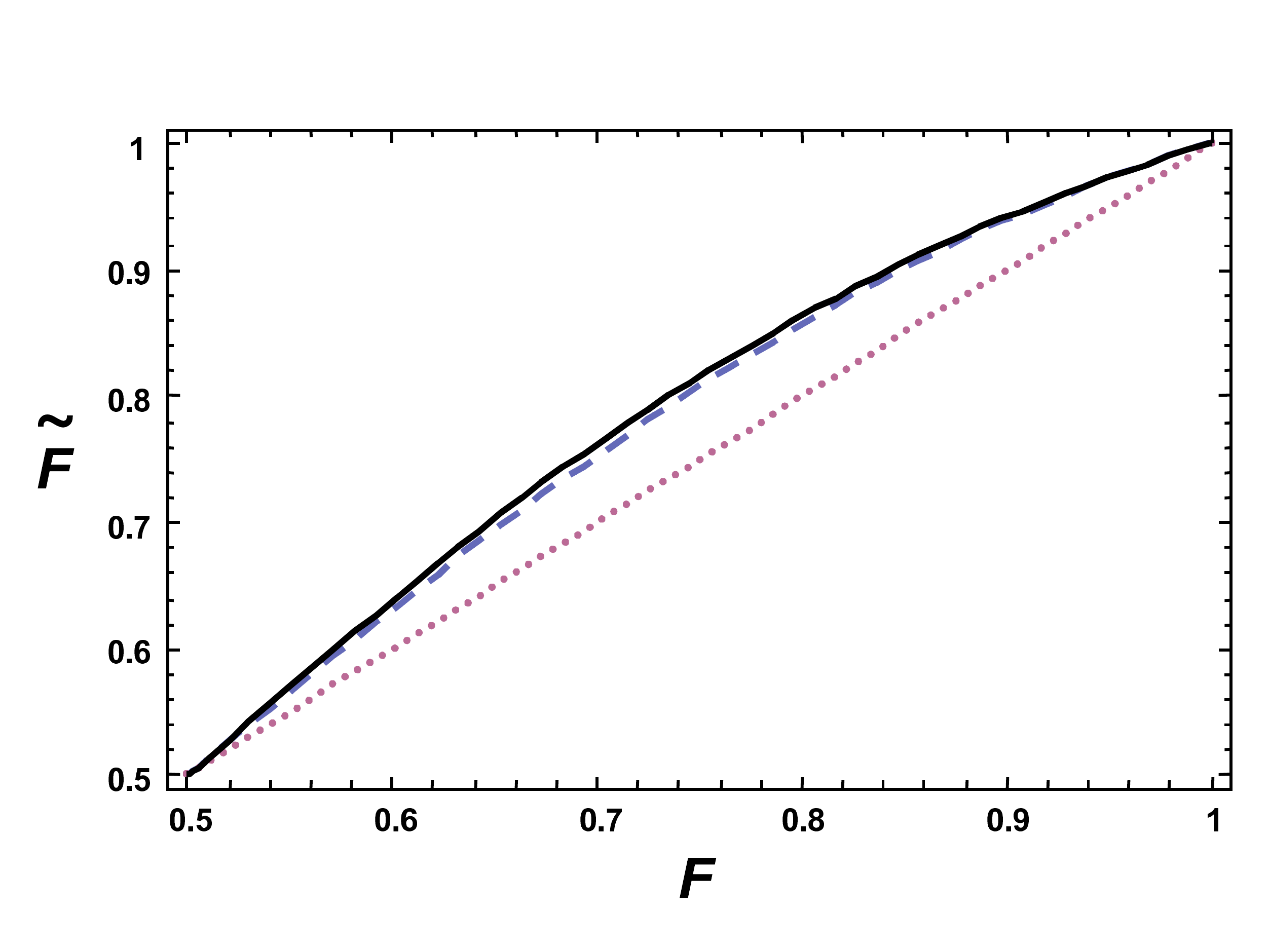}
\caption{(Color online) Output fidelity as a function of
input fidelity for the optimized protocol, where the
parameter $\theta$ is adapted to the input fidelity
($\tilde{F}_{opt}$, full), and for a simplified protocol
where $\theta=\frac{\pi}{8}$ for all input fidelities
($\tilde{F}_1$, dashed). As a reference the straight line
$\tilde{F}=F$ is also shown (dotted).}\label{fig2}}
\end{figure}

The proposed protocols are very feasible with current
technology. For example, for the simplified protocol,
Alice's linear operation $U$ simply corresponds to a beam
splitter with (amplitude) transmission coefficient $\cos
\frac{\pi}{8}$, corresponding to an intensity transmission
of $85 \%$, and Bob's operation $V$ to a beam splitter with
amplitude transmission coefficient $\cos \frac{-3\pi}{8}$,
corresponding to an intensity transmission of $15 \%$. The
optimized protocol requires beam splitters with varying
transmission depending on the input fidelity $F$. Note that
if the modes $a_1$ and $a_2$ (and analogously $b_1$ and
$b_2$) are converted to the polarization states of a single
spatial mode \cite{Chou07}, then the required generalized
beam splitters are very easy to realize combining wave
plates and polarizing beam splitters.

The protocol relies on single-photon interference and on
the bosonic character of indistinguishable photons.
Single-photon interference is equivalent to classical
interference and can be performed with extremely high
precision. The most significant experimental challenge is
likely to be the generation of highly indistinguishable
photons. Their degree of indistinguishability can be
quantified by their mode overlap, which corresponds
experimentally to the visibility $V$ of the
``Hong-Ou-Mandel dip'' \cite{HOM}, i.e. the extent to which
the two photons ``bunch'' after a 50/50 beam splitter
($V=1$ corresponds to perfect bunching). Ref.
\cite{fransondip} has reported a very impressive visibility
$V=0.994$ , which is largely sufficient for high-fidelity
purification, cf. Fig. 3. The result of Ref.
\cite{fransondip} was achieved for photons from the same
pair for a parametric down-conversion source, which is
likely to be the most promising system for initial
demonstrations of the proposed protocol. In general, Fig. 3
shows that purification is possible for dip visibilities
that should be achievable in a wide range of systems.

\begin{figure}
{\includegraphics[scale=0.32]{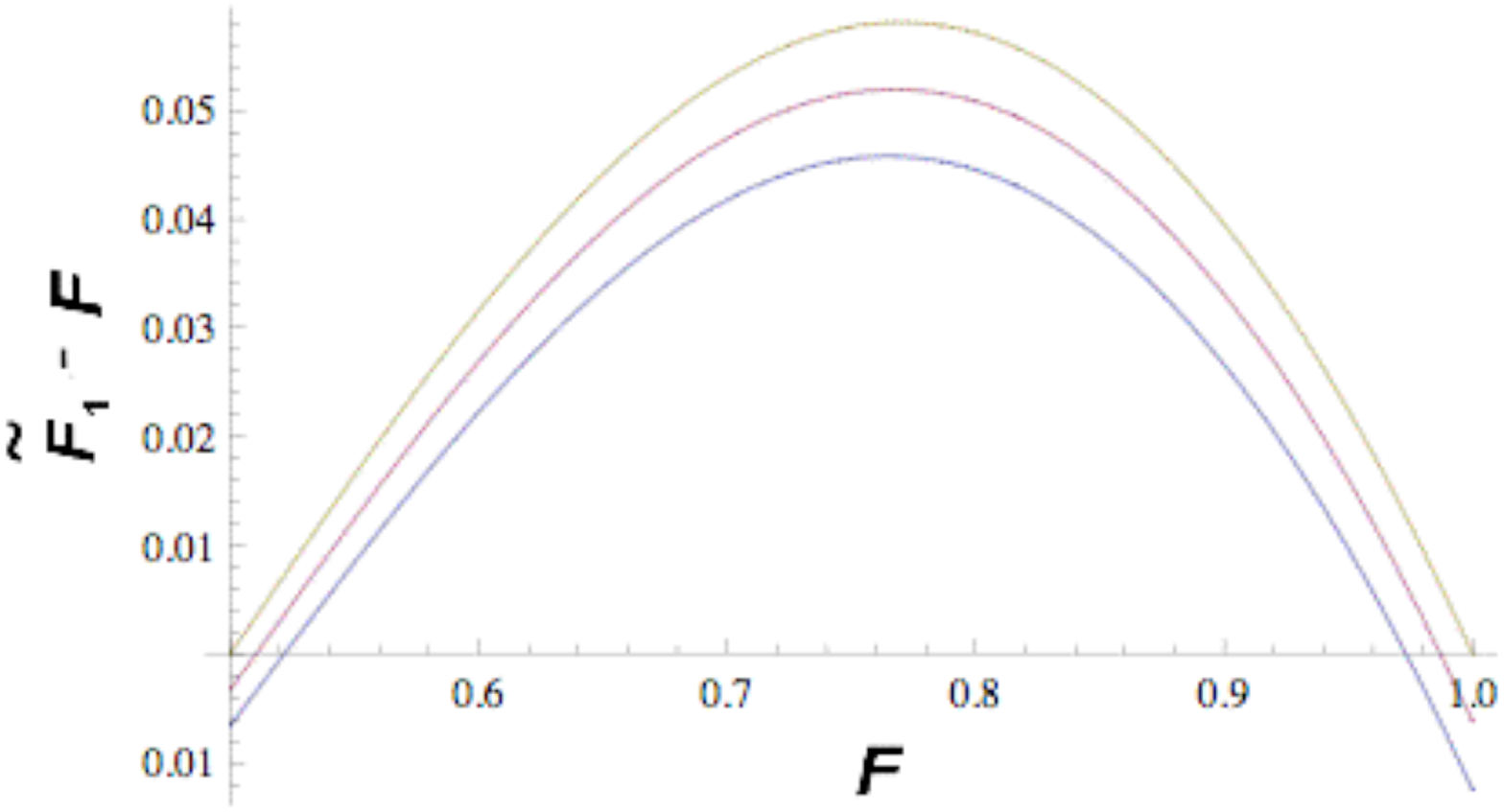}
\caption{Fidelity improvement $\tilde{F}_1-F$ as a function
of $F$ for Hong-Ou-Mandel dip visibilities $V=1, 0.99$ and
$0.98$ (top to bottom).}\label{fig3}}
\end{figure}

The protocol in its ideal form requires highly efficient
photon-number resolving single-photon detectors. We now
discuss the impact of non-perfect detection efficiency. The
non-detection of a second photon that is actually present
in mode $d_a$ or $d_b$ will lead to a vacuum component in
modes $\tilde{a}, \tilde{b}$. However, it will not reduce
the fidelity for the single-photon component of the output.
If the purification protocol is applied in the context of
the quantum repeater schemes of Refs.
\cite{SinglePhotonRepeaters}, the vacuum component leads to
a lower success probability for the subsequent entanglement
swapping steps and thus to a lower entanglement generation
rate. On the other hand, the fidelity of the generated
long-distance entanglement only depends on the fidelity of
the single-photon component (and thus is independent of the
detection efficiency), since the final measurement in these
schemes post-selects the cases where there was a photon in
the output. Note that highly efficient detectors are being
developed, e.g. Ref. \cite{lita08} has recently reported
photon-number resolving detectors with $95 \%$ efficiency
at 1556 nm, a wavelength ideally suited for long-distance
transmission in optical fibers.

The proposed scheme could be demonstrated both in purely
photonic experiments, and in experiments involving quantum
memories for photons, as an important further step towards
the realization of a quantum repeater. The first type of
experiment, in addition to the components discussed above,
just requires two sources of single photons, which could be
realized conditionally based on parametric down-conversion
\cite{PDC-SPS} or directly using quantum dots
\cite{Santori02} or single atoms \cite{Kuhn02}.

The second type of experiment would be very similar to the
experiment of Ref. \cite{Chou07}. Single-particle
entanglement would first be created between two pairs of
atomic ensembles. The stored excitations are then
reconverted into photons, combined on linear optical
elements as described above and detected. The conditions
for the successful realization of the purification scheme
proposed here, in particular indistinguishability of the
photons emitted by different ensembles, are similar as for
the experiment of Ref. \cite{Chou07}. Interesting further
experimental steps would be to re-absorb the purified
delocalized photon in a memory, and of course to increase
the distance between the two parties.

In conclusion, we have presented a very simple scheme for
the purification of single-photon entanglement which is
realizable with current technology. We have furthermore
shown that the scheme achieves the optimal fidelity for any
number of auxiliary vacuum modes. We find the simplicity of
the scheme remarkable from a conceptual point of view. It
also constitutes important progress for the implementation
of quantum repeaters based on single-particle entanglement
\cite{SinglePhotonRepeaters}.

We thank J.-D. Bancal, C. Branciard, N. Brunner, R.
Dubessy, H. de Riedmatten, and H. Zbinden for helpful
comments. This work was supported by the EU Integrated
Project {\it Qubit Applications}, the Swiss NCCR {\it
Quantum Photonics} and the French National Research Agency
(ANR) project ANR-JC05\_61454.



\begin{thebibliography}{100}
\bibitem{phase} Note that the state of Eq. (1) is
obtained by averaging the state
$\frac{1}{\sqrt{2}}(|0\rangle_A |1\rangle_B+e^{i \varphi}
|1\rangle_A|0\rangle_B)$ over $\varphi$ with a distribution
such that the mean value of the phase factor is $\langle
e^{i\varphi}\rangle=2 F-1$.
\end{thebibliography}
\end{document}